\documentclass[conference]{IEEEtran}
\IEEEoverridecommandlockouts
\usepackage{cite}
\usepackage{amsmath,amssymb,amsfonts}
\usepackage{graphicx}
\usepackage{textcomp}
\usepackage{xcolor}
\usepackage{comment}
\usepackage{algorithm}
\usepackage{algpseudocode}
\usepackage{romannum}
\usepackage{subcaption}
\def\BibTeX{{\rm B\kern-.05em{\sc i\kern-.025em b}\kern-.08em
    T\kern-.1667em\lower.7ex\hbox{E}\kern-.125emX}}

\begin{document}

\title{An Initial Access Optimization Algorithm for millimeter Wave 5G NR Networks
\thanks{Authors would like to certify that this work has not been published in
any other conference or has not been submitted for any other publication
elsewhere.
}
}

\author{A. Indika Perera, K. B. Shashika Manosha, Nandana Rajatheva, and Matti Latva-aho
        \\Centre for Wireless Communications, University of Oulu, Finland.
        \\\{indika.perera, manosha.kapuruhamybadalge, nandana.rajatheva, matti.latva-aho\}@oulu.fi}

\maketitle

\begin{abstract}

The millimeter wave (mmWave) communication uses directional antennas. Hence, achieving fine alignment of transmit and receive beams at the initial access phase is quite challenging and time-consuming.
In this paper, we provide a dynamic-weight based beam sweeping direction and synchronization signal block (SSB) allocation algorithm to optimize the cell search of the initial access in mmWave 5G NR networks. The number of SSBs transmitted in each beam sweeping direction depends on previously learned experience which is based on the number of detected UEs (user equipment)  per SSB for each sweeping direction. Overall, numerical simulation results indicate that the proposed algorithm is shown to be capable of detecting more users with a lower misdetection probability. Furthermore, it is possible to achieve the same performance with a smaller number of dynamic resource (i.e., SSB) allocation, compared to constant resource allocation.

\end{abstract}


\section{Introduction}

Fifth generation (5G) new radio (NR), the mobile communication standard presented by the 3GPP as the 3GPP release 15, introduces major improvements to existing long term evolution-advanced (LTE-A) standard \cite{NR38211}. 
To achieve the 5G goals, 3GPP has introduced an unified network architecture, with a new physical layer design that supports very high carrier frequencies (mmWaves), large frequency bandwidths, and new techniques such as massive MIMO (multiple-input and multiple-output),  and beamforming  \cite{NR38213}.

The frequency spectrum ranging 30-300 GHz (wavelengths 1-10 mm) is collectively known as millimeter wave (mmWave) bands. A large amount of continuous bandwidths available at mmWave frequencies compared to the existing sub 6 GHz radio spectrum which is fragmented and crowded, has vast potential to increase the capacity of 5G cellular wireless systems which are in gigabits per second rates \cite{7000981}. 
Moreover, due to the shorter wavelength of mmWaves, the physical size of the antenna arrays becomes small. Therefore, using antenna arrays for mobile communications and realizing directional communication with very narrow beams are possible. Thus, with narrow beams accurate beamforming operations can be performed, allowing to focus the beam in very precise spatial directions \cite{wang2018compact}. Hence, a highly directional transmission is inherent to mmWave communications, and using that high isotropic pathloss of the mmWave frequencies can be mitigated.

5G and above generations of cellular networks which planned to use mmWave communication must provide a set of mechanisms to provide the directional beam management.
Most importantly these directional links require fine alignment of the transmitter (Tx) and receiver (Rx) beams. This is a new challenge the new generation of communication systems face compared to the earlier generation of communication systems.
Initial access (IA) in a cellular communication is defined as the establishment procedure of an initial connection between an user equipment (UE) and a base station (BS) in a mobile cellular network. Initial access is a critical prerequisite for any subsequent communication between the BS and UE. In mmWave communication, initial access is further challenging due to directional communication. This is because BS and UE  are required to find the correct Tx and Rx beam pair which is aligned with each other, without the location information about BS and UE.

In \cite{giordani2016comparative}, authors compare two initial access algorithms named exhaustive and iterative, both consist of a complete scan of the whole $360^{\circ}$ angular space by the BS and the UEs in search for the best communication channel between them, according to fixed codebooks. 
Exhaustive search is the basic repetition method available. It is a brute force sequential beam searching method which can be applied to both BS and UE. 
Iterative search is a two-stage beam scanning procedure where it uses two types of  BS beams for beam sweeping the BS coverage area. Simulations have shown that exhaustive search has higher discovery delay and iterative search has higher misdetetction probability.
The hybrid search algorithm presented in  \cite{wei2017exhaustive}, is similar to the iterative search algorithm with some improvements to further reduce the discovery delay.
The algorithm proposed in \cite{parada2017cell}, accounts for the scan of the whole space like the aforementioned exhaustive search, however, in this case, the order in which the sectors are scanned is based on the previous experience. Studies on an initial access technique with an \textit{optimized number of synchronization signal block (SSB) allocation per sweeping direction, based on user distribution}, is not found in the existing literature.

In order to overcome the high path loss and to cater for the high traffic hot-spots, small cells are proposed for mmWave communication. Hence, most of these mmWave small cells will be in the urban areas \cite{7343692}. Since these are in urban areas, a large number of obstacles such as buildings, vehicles and furniture will be within the coverage area of these small cells. This will severely obstruct the mmWave line of sight (LOS) communication between the BS and the UE. Most of these small cells are targeting office environments and public places where a large number of users gather. Consequently, users are scattered around the BS as clusters, and due to that, some directions may have higher user density while some directions may have a lower or no user density at all. Hence, it is quite obvious that users served by a small cell BS may not be distributed uniformly around the BS.

Therefore, cell search procedure for this kind of scenario requires some intelligence at the BS. Thus, providing knowledge about user distribution, BS is able to steer its beams through a known populated area for UEs instead of wasting time and energy transmitting towards an obstacle or less populated directions. 
Nevertheless, the BSs considered in current research arena, are not dynamically deciding which areas to cover or not based on user distribution. However, it is highly beneficial for BS to have some dynamic update about the UE distribution around it.

In this paper, we present a new dynamic-weight based algorithm for \textit{calculating the optimized SSB allocation of each beam sweeping direction} 
for initial access in mmWave 5G NR networks
\cite {perera2019initialThesis}\footnote{This paper is based on the research findings of the first author's master's thesis \cite {perera2019initialThesis}.},\cite{perera2019initial}\footnote{A pre-conference version of this paper is uploaded to \cite{perera2019initial}.}.
Optimized SSB allocation is based on user distribution, which is calculated using the number of detected UEs
per SSB for each sweeping direction. Our numerical simulation results demonstrate that the proposed algorithm has higher detection probability and optimized resource usage compared to existing initial access algorithms.

The rest of the paper is organized as follows. In Section \ref{s2}, we introduce the 5G NR synchronization signal structure and our system setup, while in Section \ref{s3} we describe the proposed algorithm in detail. In Section \ref{s4} we evaluate the performance of the algorithm through simulation results. Finally, in Section \ref{s5}, we summarize our major findings.

\section{System Model}
\label{s2}

 \subsection{Synchronization signal structure in 5G NR}

In 5G NR, a synchronization signal (SS) consists of two main signals known as primary synchronization signal (PSS) and the secondary synchronization signal (SSS). These synchronization signals, together with the physical broadcast channel (PBCH), are referred to as a synchronization signal block (SSB) or SS block in 5G NR definitions. SSB is a group of 4 OFDM symbols in time and 240 subcarriers in frequency (i.e., 20 resource blocks) in NR time-frequency grid \cite{ETSITS1383001}.
There are several SSBs in one NR radio frame. The number of SSBs and the time domain position of these SSBs depend on NR numerology. This is a significant difference compared to the LTE, where its number of synchronization signals and time domain positions are not changing.

One important feature introduced in NR, related to initial access beam sweeping is the possibility to apply beam-sweeping for SSB transmission, which means the possibility to transmit synchronization signal blocks in different beams in a time-multiplexed fashion \cite{dahlman20185g}. The set of SSBs within a beam-sweep is referred to as \textit{SS burst set}. 
The periodicity of the SS burst set is flexible with a minimum period of 5 ms and a maximum period of 160 ms.

The number of SSBs needed to transmit within a radio frame is not fixed. There is the flexibility of transmitting a single SSB up to the possible maximum number of SSBs within a SS burst set. Depending on the number of beams needed for the beam sweep, one can decide on the number of transmitting SSBs per SS burst set \cite{dahlman20185g}.  This is an important feature in NR that we used in developing our proposed initial access algorithm.

\subsection{System setup}
\label{sysSetup}

We consider a scenario of an urban small cell with a single 5G BS. Considering the radius of typical 5G mmWave small cell, 150 m cell radius is selected. We assume a static user deployment where no users are moving during the simulation time period. Within the cell area, 40 users are distributed uniformly around the BS with different radii between 5 m and 150 m. UEs are assumed to have random antenna orientation range from 0-$360^{\circ}$ with respect to the reference direction. NLOS Rayleigh channel at 28 GHz is considered between BS and UE.

We use 5G NR numerology 3, with 120 kHz subcarrier spacing as the radio frame structure. For frequencies higher than 6 GHz, 120 kHz subcarrier spacing is designed
and with this frame structure maximum of 64 SSBs can be configured for a single radio frame.

\begin{figure}[htbp]
  \begin{subfigure}[h]{0.24\textwidth}
     \includegraphics[trim={4cm 1.5cm 3.7cm 1.15cm},clip,scale=0.6]{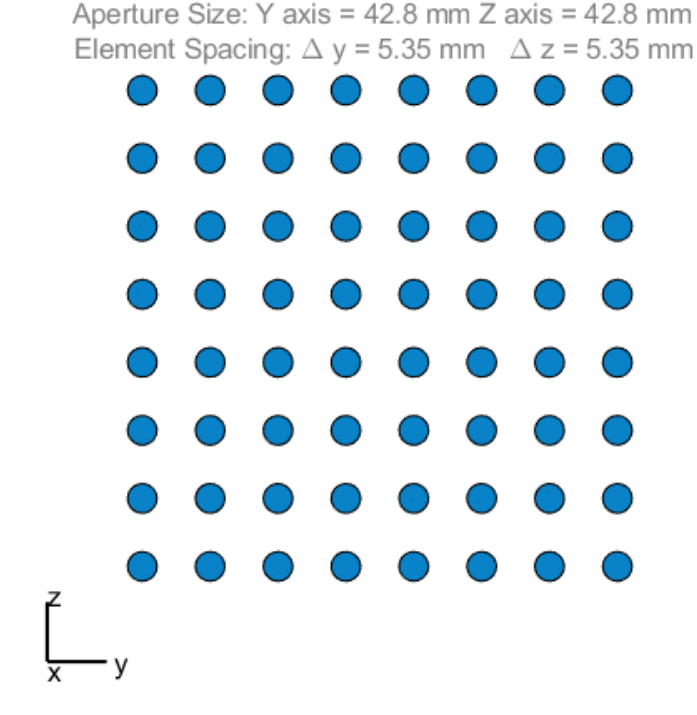}
     \caption{}
     \label{88array1}
  \end{subfigure}
  \hfill
  \begin{subfigure}[h]{0.24\textwidth}
    \includegraphics[trim={0cm 0cm 0cm 0cm},clip,scale=0.45]{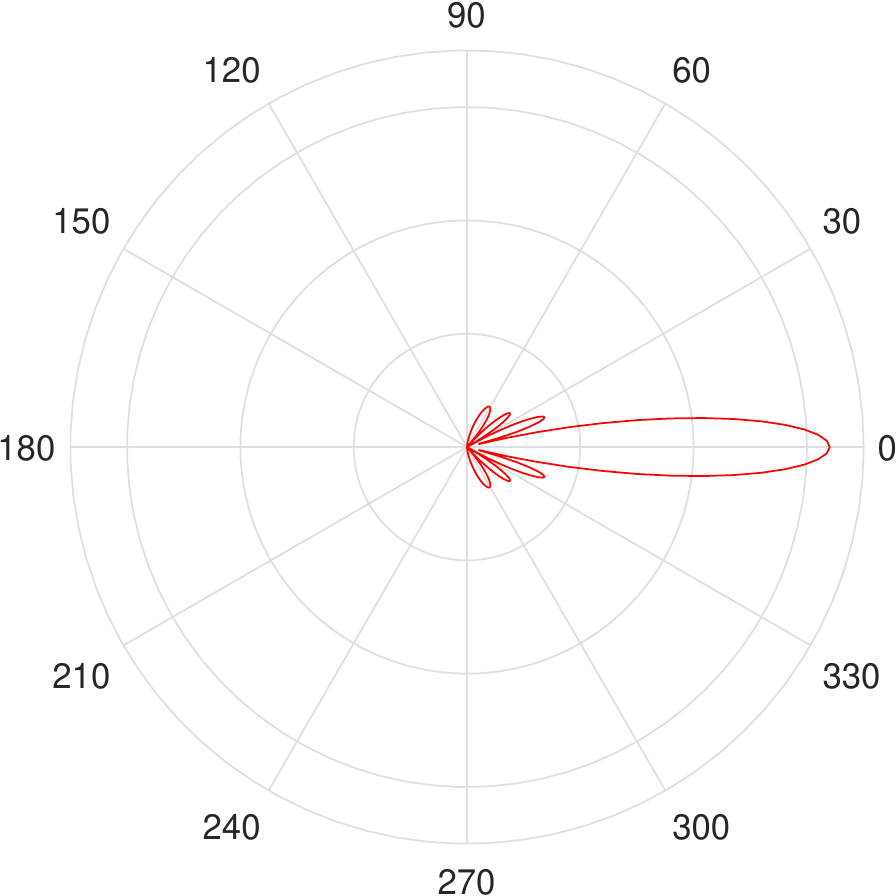} 

 \caption{}
  \label{88array2}
   \end{subfigure}
   
  \caption{Antenna geometry and 2D beam pattern of $8\times8$ UPA}
    \label{88array}
\end{figure}

At the BS side $8\times8$ uniform planner array (UPA) with $\lambda/2$ antenna element spacing (as shown in Fig. \ref{88array1}) is used. For simulating equal beam pattern, 4 sectors each has an $8\times8$ UPA are mounted on the BS towards the $45^{\circ}$, $135^{\circ}$, $225^{\circ}$ and $315^{\circ}$ directions.
Further, we assume that the UE is equipped with $2\times2$ UPA panel. Both BS and the UE antenna elevation angles are considered as zero for simulating 2D beam pattern. 2D beam pattern used as the BS narrow beam is shown in Fig. \ref{88array2} below.

There are two types of beam patterns used in BS; BS wide beams (WB) with lower beamforming gain and BS narrow beams (NB) with higher beamforming gain.
We consider four wide beams with an angular region of $90^{\circ}$ in azimuth at the UE. At the BS, four wide beams same as UE wide beams, and 16 narrow beams with an angular region of $22.5^{\circ}$  are considered.

\section{Proposed Algorithm}
\label{s3}

In this section, a new dynamic-weight based algorithm for calculating the \textit{optimized SSB allocation for each beam sweeping direction} in the initial access procedure is presented.
Within a beam sweep cycle, the number of SSBs allocated for each beam sweeping direction depends on the user distribution learned previously. This knowledge (i.e., weights) is based on the number of detected UEs per SSB for each sweeping direction. Thus, by providing some knowledge about the UE distribution around the BS, the delay in detecting UEs during the cell search procedure will decrease, and the number of UEs found during the beam scan will increase.

\subsection{Proposed initial access algorithm}

\begin{figure}[htbp]
\centerline{\includegraphics[trim={2.8cm 10.1cm 6.8cm 1.9cm},clip,scale=0.73]{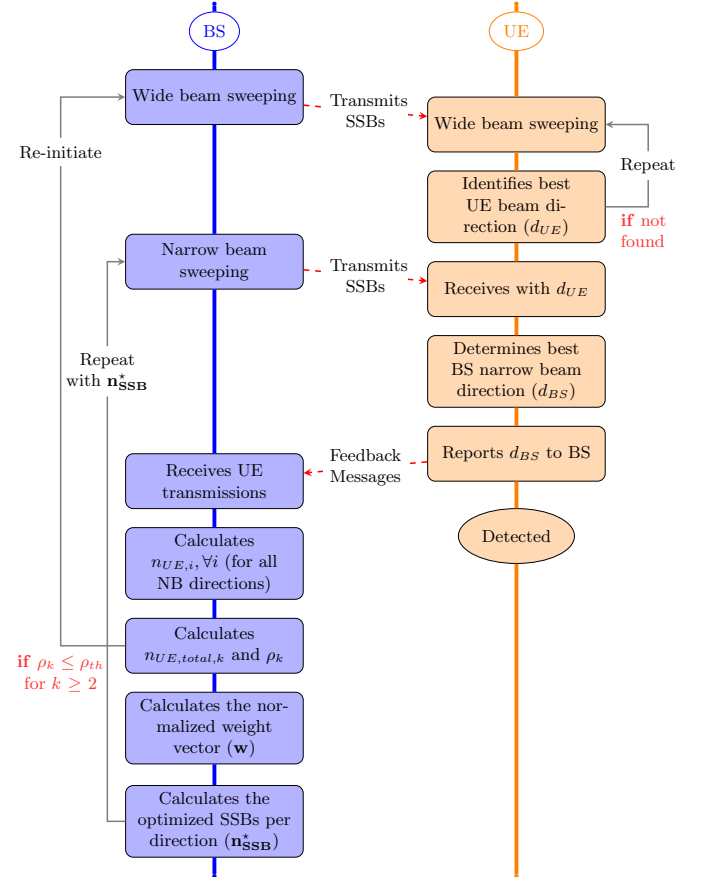}}
\caption{Flow diagram of the proposed IA algorithm.}
\label{fig1}
\end{figure}

 The flow diagram of the proposed initial access algorithm is shown in Fig. \ref{fig1}. This algorithm can be mainly divided into two parts: the BS side algorithm (see left-hand side of Fig. \ref{fig1}) which runs in the serving BS and UE side algorithm (see right-hand side of Fig. \ref{fig1}) which runs in each of the UEs trying to conduct the initial access. Both of these algorithms are running in parallel but may start in two different time instances based on the time UE starts its initial access procedure. Note that the dashed arrows represent the actions which BS and UE interact. Procedures of these two algorithms are separately described in algorithm  \ref{BS algo}  for BS and  algorithm \ref{UE algo} for UE.

\begin{algorithm}
  \caption{BS Algorithm}\label{BS algo}
  \begin{algorithmic}[1]
  	
  	\State  BS transmits SSBs using wide beams
  	\State  BS transmits SSBs using narrow beams
 	\State  BS receives UE feedback messages
 	\State  BS calculates the detected number of UEs for each narrow beam direction ($n_{UE,i}$, $\forall i$)
 	\State  BS calculates detected UEs up to $k$th beam sweep cycle ($n_{UE,total,k}$) and detection ratio ($\rho_{k}$) for $k \geq 2$
        \newline\textbf{if } $\rho_{k} \leq \rho_{th} $ \textbf{then} move to step 1 \textbf{else} move to next step
 	\State  BS calculates the normalize weight vector ($\bf{w}$)
 	\State  BS calculates the optimized SSBs per direction vector ($\bf{n^{\star}_{SSB}} $)
 	\State Repeat step 2 onwards with optimal SSB allocation $\bf{n^{\star}_{SSB}} $ and new beam sweeping order
  	
  \end{algorithmic}
\end{algorithm}

\begin{algorithm}
  \caption{UE Algorithm}\label{UE algo}
  \begin{algorithmic}[1]

  	\State  UE receives SSBs using wide beams 
  	\State UE identifies best UE beam direction ($d_{UE}$)
            \newline \textbf{if} $d_{UE}$ found \textbf{then} move to next step \textbf{else} move to step 1 
 	\State  UE starts to receive in $d_{UE}$ direction
 	\State  UE determines best BS narrow beam direction ($d_{BS}$)
 	  \newline \textbf{if } $SNR_{d_{BS}} \geq SNR_{th}$ 
 	        \textbf{then} move to next step  \textbf{else} keep on searching
 	\State UE feeds back the $d_{BS}$ to BS
 	\State UE is detected. Proceed to RA
  	 	    	  
  \end{algorithmic}
\end{algorithm}

The flow diagram and algorithms contain the same steps, hence only the steps of the BS and UE algorithms are explained. Since these two algorithms are executed at the same time, they are described according to the sequential order they occur.

\subsection{Explanation of IA algorithm}

Let us denote the total number of wide beam sweeping directions in BS as $D_{BS,wb}$, the total number of narrow beam sweeping directions in BS as $D_{BS,nb}$, the total number of beam sweeping directions in UE as $D_{UE}$ and the total Number of SSBs in a single radio frame (i.e., SS burst set size) as $n_{SSB,total}$.

As the first step of the BS initial access algorithm, BS starts to beam sweep with its wide beams.
Consecutive SSBs in a radio frame is transmitted in consecutive beam sweeping directions (one SSB per a direction), and this process is repeated until $\frac{1}{2}\times n_{SSB,total} $ is transmitted. 

The UEs around the BS which are intended to join the network starts to run the UE side initial access algorithm. They need to scan and find the correct UE beam to communicate with the BS. As the first step of the UE algorithm, UE starts to beam sweep  $360^{\circ}$ and receives the synchronization signals (i.e., SSBs) from the BS.

As in step 2 of the UE algorithm, when UE receives 
synchronization signals from the BS to cover all UE beam directions ($D_{UE}$), UE identifies the best UE beam direction denoted by $d_{UE}$. It is the UE beam direction with the highest received SNR. If UE is unable to find the best UE beam direction due to insufficient number of received signals or low received signal power, it continues to beam sweep further. This process is continued until the UE has discovered its best beam direction.

Once the best UE beam direction ($d_{UE}$) is found, UE beamforms to that direction to receive future transmissions from the BS. Thus, the UE  stops beam sweeping and starts receiving the transmission signals from the BS, as shown in step 3 of the UE algorithm. 

As step 2 of the BS algorithm, BS starts to transmit the remaining $\frac{1}{2}\times n_{SSB,total} $ number of SSBs of the radio frame using the BS narrow beams. This sweeping is also done as earlier by transmitting consecutive SSBs in consecutive directions. When the BS narrow beam sweeping is happening, UE can be in either of the following two states.
(\romannum{1}) Receiving these BS narrow beams by beamforming to  $d_{UE}$. 
(\romannum{2}) Continues to beam sweep with UE beams for identifying $d_{UE}$. 

After the reception of synchronization signal via BS narrow beams, UE tries to determine the best BS narrow beam direction ($d_{BS}$) as step 4 of the UE algorithm. Thus, the UE calculates the SNR of all the received signals via narrow beams when UE beamformed to $d_{UE}$. Then the direction of the highest received SNR value ($SNR_{d_{BS}}$) is identified as the best BS narrow beam direction $d_{BS}$, if  $SNR_{d_{BS}}$ satisfies the inequality $SNR_{d_{BS}} \geq SNR_{th},$ where  $SNR_{th}$  is the predefined SNR threshold value required to establish a reliable  communication link between BS and UE. If the above inequality is not satisfied, then this link is not in a condition for reliable communication. Therefore, the UE keeps on searching for a BS narrow beam direction or other possible BS that can satisfy this condition.

As the step 3 of the BS algorithm, once BS narrow beam sweeping is completed, BS beam sweeps to every narrow beam direction to receive the UE messages. Thus, the BS starts to receive the UE feedback messages. At this time if there are any UEs which were able to determine the $d_{BS}$, those transmit the identified direction to the BS as shown in step 5 of the UE algorithm.

After this step, UE knows its best beam and BS knows its best narrow beam to communicate with each other. Therefore, UE is considered as a detected UE to that BS. Hence, the UE side initial access algorithm stops here, and UE proceeds to the random access (RA) procedures.

Once the BS received these UE feedback messages, as shown in the BS algorithm step 4, BS calculates the number of UEs detected for each narrow beam sweep direction $n_{UE,i}$ for all $ i=1, \ldots,D_{BS,nb} $.

Then as BS algorithm step 5, BS calculates the total number of UEs detected 
up to the complete beam sweep cycle $k$,  $n_{UE,total,k}$. At this stage, we define detection ratio $\rho_{k}$ for $k$th sweep cycle as, $\rho_{k} = {n_{UE,total,k}}/{n_{UE,total,k-1}} $
for all the $k \geq 2$ sweep cycles.
As step 5 of the BS algorithm,  BS calculates the $\rho_{k} $ and determines whether to proceed forward to optimization procedures or re-initiates the beam sweep process based on the inequality $\rho_{k} \leq \rho_{th},$
where $ \rho_{th}$  is predefined threshold ratio value (default value is one) set to satisfy the user detection capability. If the  calculated $\rho_{k} $ value satisfies the said inequality  then the beam sweeping is re-initiated and algorithm returns to step 1 of BS algorithm. Otherwise, BS proceeds to the optimization procedures as explained below.

The basic idea behind the introduced optimization procedure is to transmit synchronization signal blocks (SSBs) according to an expected number of users in that beam sweeping direction. 
It is assumed when a higher number of SSBs transmitted,  UE within that area has a higher probability of receiving and decoding those SSBs. Hence, a higher number of UEs could be detected from that direction. 
Also, for situations where there are no users in a given direction, transmitting SSBs repetitively for that direction is a waste of resources. 
Therefore, the optimal solution is to transmit a single SSB for that direction during a single radio frame.

An underlying problem when deciding the number of SSBs required for a given direction is, the basis we select to decide it. In the proposed algorithm, number of users detected per a SSB is used as the basis for deciding it. In BS algorithm step 6,  BS calculates the  weight vector $\bf{\overline{w}} \in \mathbb{R_{+}}^{D_{BS,nb}} $ with elements 
\begin{equation}
  \overline{w_{i}} = {n_{UE,i}}/{n_{SSB,i}} , \forall  i , \label{eq6}
\end{equation}where $n_{SSB,i}$ is the number of SSBs transmitted to $i$th narrow beam sweep direction. Then the BS calculates the normalized weights vector $ \bf{w} \in \mathbb{R_{+}}^{D_{BS,nb}} $ as
 \begin{equation}
 \bf{w} = {\bf{\overline{w}}}/{\Vert \bf{\overline{w}}\Vert_{1}}.
 \label{eq7}
 \end{equation} 
The normalized weight vector defined above provides us with information about the currently detected user distribution among the transmitting narrow bream directions. We can use it to predict the number of users that will be discovered in the next beam sweep cycle.

Number of SSBs transmit per direction vector $\mathbf{n_{SSB}}$, is a vector of size  ${1\times D_{BS,nb}} $  defined as
 $\mathbf{n_{SSB}} = [n_{SSB,1},\ldots, n_{SSB,D_{BS,nb}}]$,
which contains the number of SSBs need to transmit in each of the narrow beam sweep direction as elements of the vector. By using $\bf{n_{SSB}}$ and the calculated weights in \eqref{eq7} above, the number of users that will discover in the next beam sweep is predicted as
$\bf{w}\times\bf{n^{T}_{SSB}}.$

Then as step 7 of the BS algorithm, BS calculates the optimized number of SSBs  ($\bf{n^{\star}_{SSB}}$)  required to transmit towards each narrow beam sweep direction in the next beam sweep cycle.
It is calculated such that the number of users which will be discovered in the next beam sweep is maximized subject to following constraints,
(\romannum{1}) Total number of SSBs that could be allocated within a single radio frame should not exceed the total allowable number of SSBs in a single radio frame ($n_{SSB,total} $) as presented in \eqref{eq9a}.
(\romannum{2})  At least one SSB is allocated to each direction. Hence, number of SSBs transmit per direction should be a positive integer as presented in \eqref{eq9b}. The integer optimization problem described above can be represented as \begin{subequations}
\label{eq9}
\begin{align}
\mbox{maximize} &\quad  \bf{w}\times\bf{n^{T}_{SSB}} \nonumber
\\
\mbox{subject to} &\quad  \sum_{i=1}^{D_{BS,nb}}n_{SSB,i}\leq n_{SSB,total}
\label{eq9a}
\\
 &\quad 
n_{SSB,i} \in \mathbb{Z}_{+} , \forall i
\label{eq9b}
\end{align}
\end{subequations} where optimization variable is  $\mathbf{n_{SSB}} \in \mathbb{Z_{+}}^{D_{BS,nb}}$. 

As the last step, calculated optimized vector using the integer optimization problem above is used to allocate the number of SSBs transmitted per each narrow beam sweep direction in the next beam sweep cycle. Then the narrow beam sweep is conducted again using the total number of available SSBs in a single radio frame ($n_{SSB,total} $). Sweeping order is not sequential this time, and it is arranged such that the highest number of SSBs allocated direction is swept first and so on. Moreover, consecutive  $n^{\star}_{SSB,i}$ number of SSBs are allocated for $i$th beam sweep direction. When compared this to the earlier narrow beam sweep cycle, earlier one was conducted in a round-robin manner. 
This concludes the proposed dynamic weight-based optimize beam sweeping direction and SSB allocation algorithm described above.

\section{Simulations and Results }
\label{s4}

We simulate the proposed optimized initial access technique using system setup mentioned in section \ref{sysSetup} above.
Apart from those mentioned earlier, other simulation parameters used for simulations are shown in Table \ref{tab1}.

\begin{table}[htbp]
\caption{Simulation parameters}
\begin{center}
\begin{tabular}{|l|c|}
\hline
\textbf{Description}&{\textbf{Value}} \\
\hline
Carrier frequency & 28 GHz \\
\hline
Channel bandwidth & 1 GHz  \\
\hline
BS Tx power & 30 dBm \\
\hline
UE Tx power & 27 dBm \\
\hline
Noise figure (NF) & 6 dB \\
\hline
Thermal noise density ($N_0$) & -174 dBm/Hz \\
\hline
\end{tabular}
\label{tab1}
\end{center}
\end{table}

\subsection{Probability of misdetection}

Probability of misdetection (PMD) is defined as the probability that a UE within a cell is not detected due to the perceived SNR at the UE is below a given SNR threshold level ($SNR_{th}$). This is a crucial performance indicator to measure the correctness of an initial access procedure. 

\begin{figure}[htbp]
\centerline{\includegraphics[trim={0.5cm 0cm 1cm 0.6cm},clip,width=0.5\textwidth,keepaspectratio]{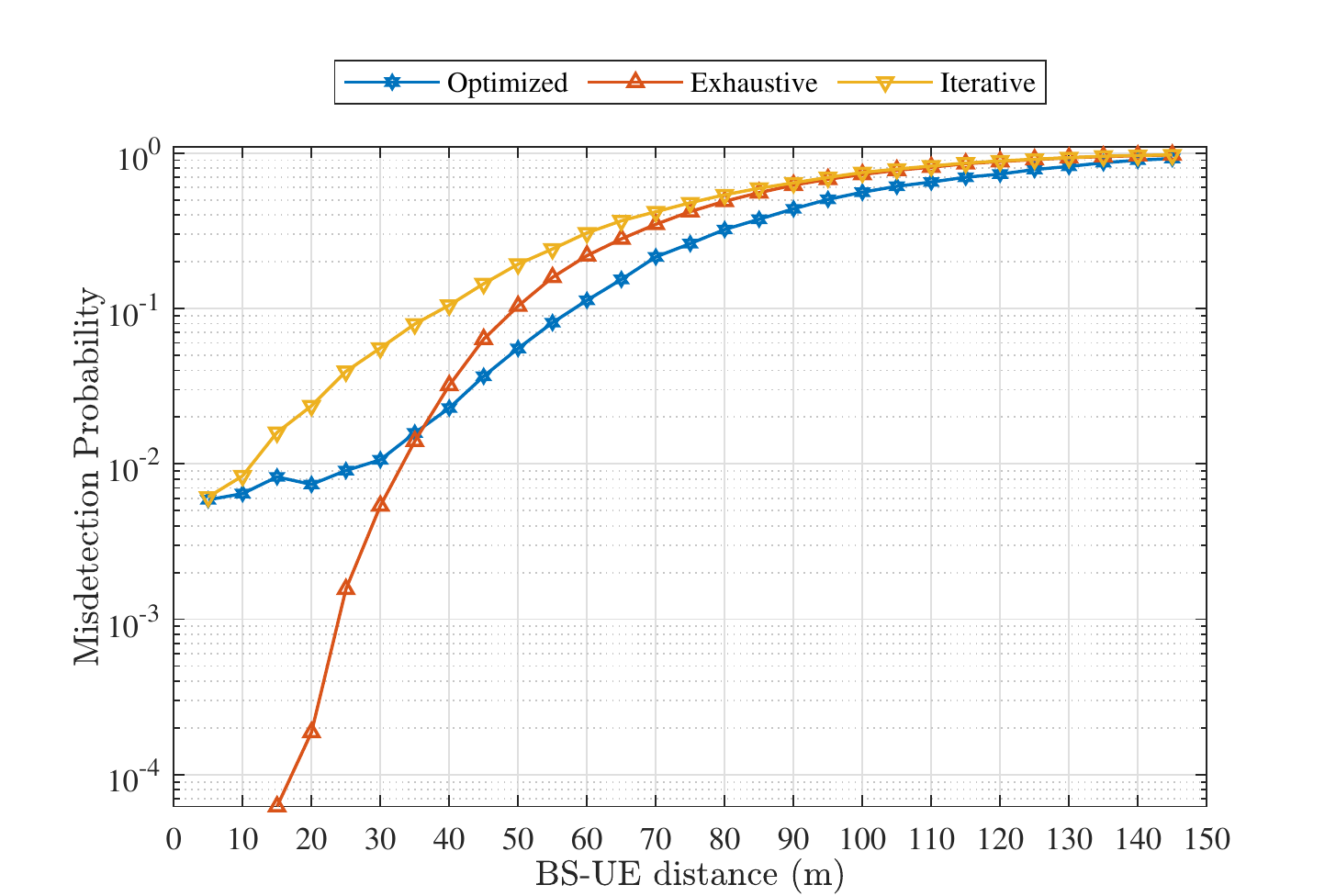}} 
\caption{PMD at SNR threshold of 0 dB.}
\label{Rf_PMD5}
\end{figure}

Simulations are done for BS to UE distance range of 5 m to 150 m.
Furthermore, the same simulation is conducted for the exhaustive and iterative initial access techniques and plotted on the same graph to compare the PMD performance of the proposed initial access technique. These two algorithms are selected since they provide the basics for initial access searching techniques and widely used for comparing the other initial access techniques. PMD at the SNR threshold of 0 dB of the proposed (optimized), exhaustive and iterative initial access techniques are shown in Fig. \ref{Rf_PMD5}.

Results in Fig. \ref{Rf_PMD5} show that the misdetection probability of all the compared schemes is increasing as the BS to UE distance increases. However, the proposed technique shows some improvement compared to the other two techniques, especially at the higher BS-UE  distances. This is mainly due to the usage of the knowledge of the historical user location for identifying the users in a given coverage range. Therefore, the PMD gain is the advantage that is achieved via optimized beam sweeping introduced in this novel algorithm.

It is worthwhile to note that at lower distances, proposed algorithm shows a bit worse performance compared to the exhaustive algorithm. The crossover point of these two PMD graphs changes with SNR threshold value which is used for detecting users. This worse performance could be the result due to the usage of wide beams to beam sweep at the first stage of the algorithm. It could be easily seen that at lower distances newly proposed optimized algorithm and iterative algorithm both show same kind of behaviour.  Since both algorithms used wide beams in the first scan, there is a higher probability that some users in the range may not be detected due to the lower beamforming gain used on those wide beams. Hence, at these distances, the exhaustive search performs well, compared to the other two algorithms. Moreover, almost all of the time iterative search demonstrates inferior performance compared to the other two algorithms.

\subsection{User detectability}

\begin{figure}[htbp]
\centerline{\includegraphics[trim={0.5cm 0cm 1cm 0.6cm},clip,width=0.5\textwidth,keepaspectratio]{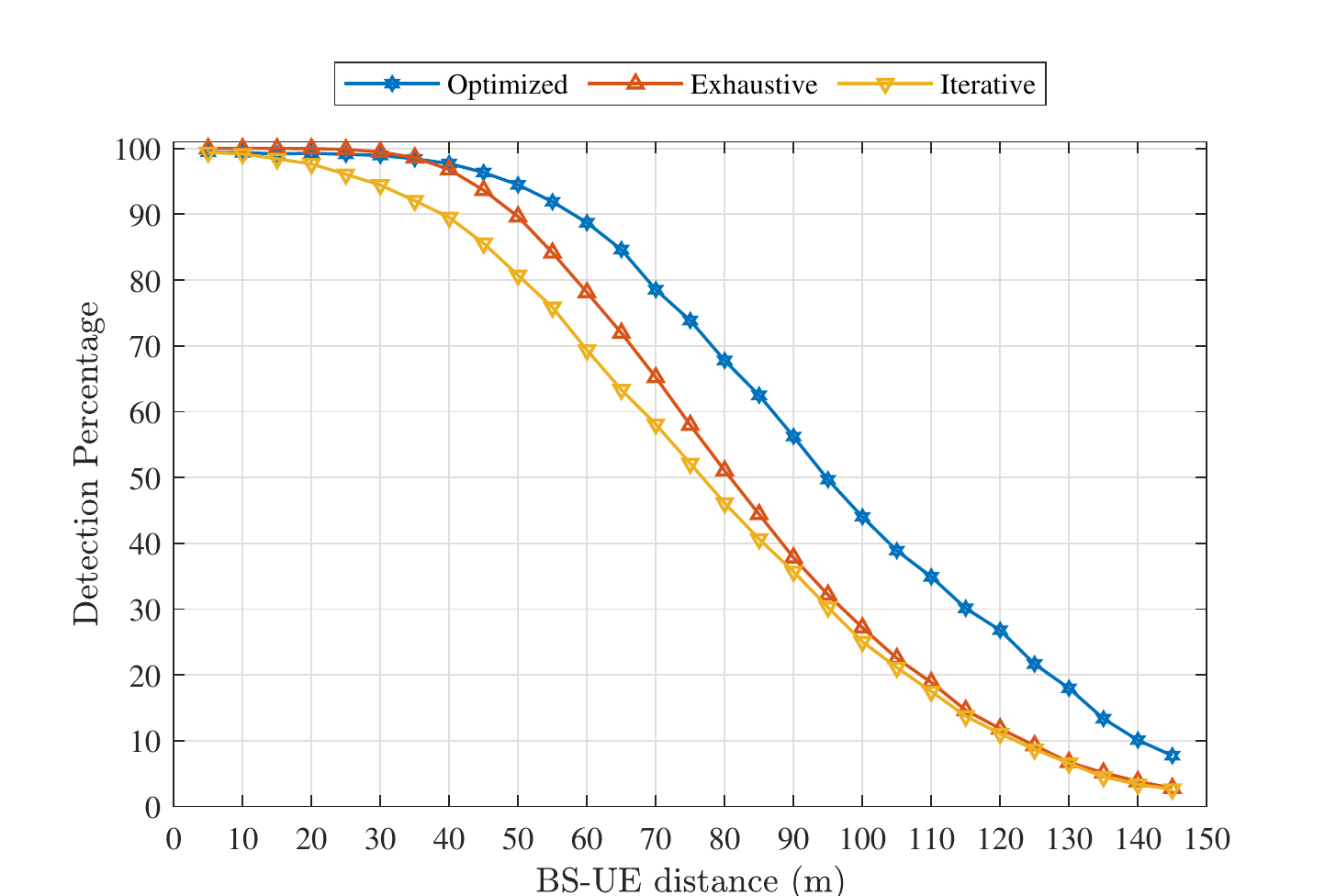}}
\caption{User detection percentage at SNR threshold of 0 dB.}
\label{Rf_Detect0}
\end{figure}

According to the same method explained earlier, detection percentage of the proposed, exhaustive and iterative methods is plotted against the BS and UE distance for a range of 5 m to 150 m 
as shown in Fig. \ref{Rf_Detect0}.

As can be seen from the Fig. \ref{Rf_Detect0}, the proposed algorithm shows a superior detection capability, especially at the higher distances from the BS compared to the other two initial access techniques. Therefore, this algorithm could provide a good detection capability for the cell-edge users who are at higher distances from BS. Even though it has a bit lower performance compared to exhaustive method at lower distances, optimized beam sweeping capability of the proposed algorithm has overcome it and improves the detectability at the higher distances.

\subsection{Resource optimization}

\begin{figure}[htbp]
\centerline{\includegraphics[trim={0.5cm 0cm 1cm 0.6cm},clip,width=0.5\textwidth,keepaspectratio]{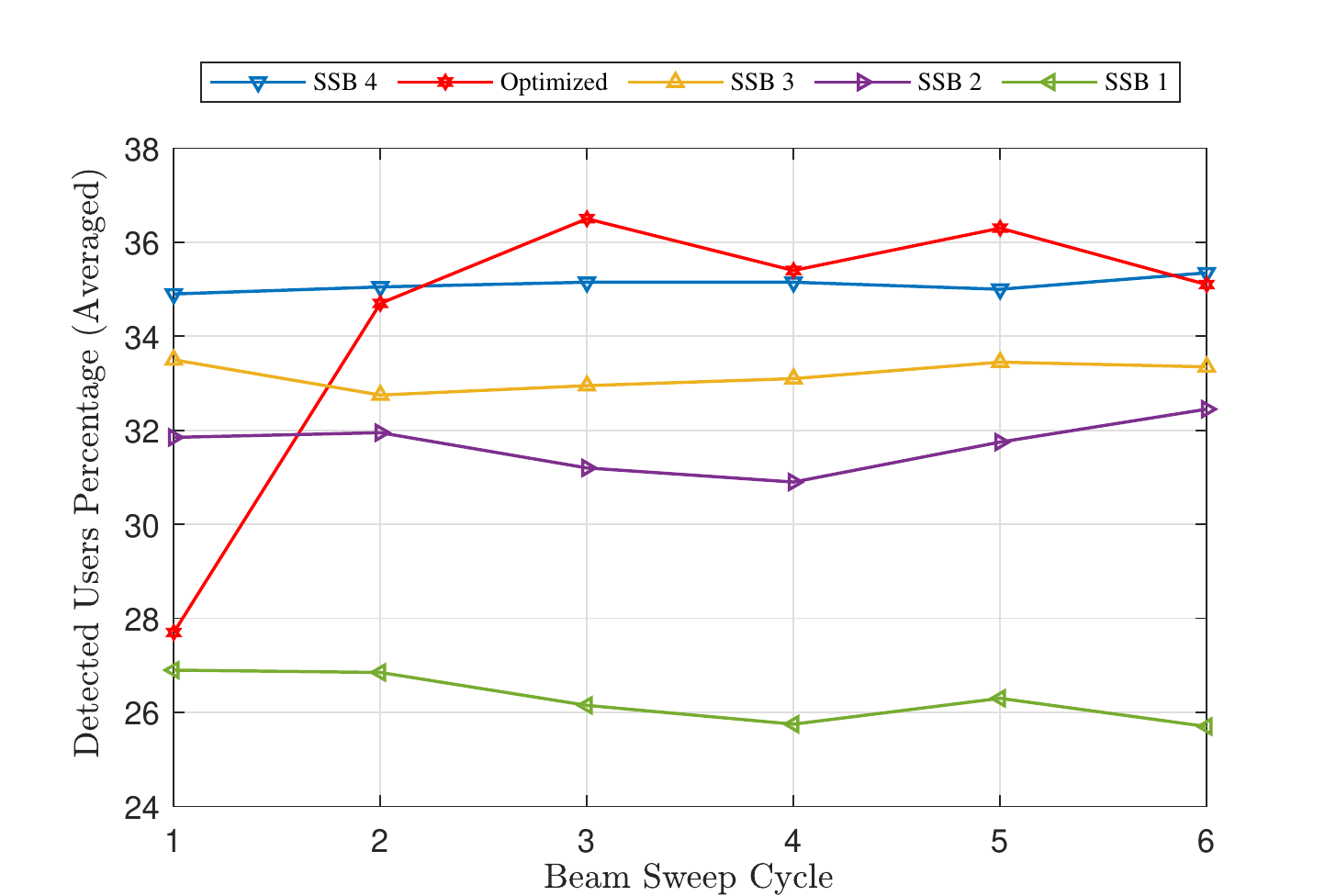}}
\caption{Averaged user detection percentage at each sweep cycle when SNR threshold of 5 dB.}
\label{Rf_DetectedUsers+5}
\end{figure}

Several simulations are conducted to evaluate the performance impact of the optimized and constant resource allocations for beam sweeping. In order to do so optimized number of SSBs per each beam sweep direction vs constant number of SSBs per each beam sweep direction is simulated. Constant SSB allocation implies that in one beam sweep cycle, every beam sweep direction is allocated an equal number of SSBs. For the considered 120 kHz NR subcarrier spacing configuration, a maximum of 64 SSBs could be allocated to whole 16 beam sweep directions we simulate. Thus, equally 1, 2, 3 or 4 SSBs per each sweep direction. Average detected user percentage at each sweep cycle against the beam sweep cycle is plotted for these allowable constant allocations and the optimized SSB allocation, is shown in Fig. \ref{Rf_DetectedUsers+5}.

From the graph in Fig. \ref{Rf_DetectedUsers+5}, it is seen that the optimized algorithm has performed better than the constant SSB allocations. Since the considered SNR threshold is bit high, a constant allocation is not capable of detecting more users due to limited constant resources in each direction irrespective of user density. Although on the other hand, optimized sweeping with allocating an optimized number of SSBs based on the expected number of users or simply the user density would be benefited in detecting more users. Hence, the optimized sweeping proposed in this new algorithm is capable of detecting more users, especially at lower SNR values or at higher distances.

\section{Conclusion}
\label{s5}

In this paper, we presented a dynamic weight-based beam sweeping direction and SSB allocation algorithm to optimize the cell search of the initial access in mmWave 5G NR networks. In the proposed algorithm, number of SSBs allocated for each beam sweeping direction depends on the user distribution learned previously. Results of the numerical simulations demonstrated that using this optimized initial access algorithm, more users could be detected with a lower misdetection probability. Also, it is possible to achieve the same performance with a lower number of dynamically allocating resources (i.e., SSBs) compared to permanently allocating a constant number of resources. In the future, this research can be extended to moving users in a multi BS  scenario where few modifications for the algorithm may be required.



\bibliographystyle{IEEEbib}
\bibliography{main}

\begin{thebibliography}{10}

\bibitem{NR38211}
{3GPP},
\newblock ``{Physical Channels and Modulation Release 15},''
\newblock Technical Specification (TS) 38.211 v15.4.0, {3rd Generation
  Partnership Project (3GPP)}, 07 2018.

\bibitem{NR38213}
{3GPP},
\newblock ``{Physical Layer Procedures for Control Release 15},''
\newblock Technical Specification (TS) 38.213 v15.5.0, {3rd Generation
  Partnership Project (3GPP)}, 05 2018.

\bibitem{7000981}
L.~{Wei}, R.~Q. {Hu}, Y.~{Qian}, and G.~{Wu},
\newblock ``{Key elements to enable millimeter wave communications for 5G
  wireless systems},''
\newblock {\em IEEE Wireless Communications}, vol. 21, no. 6, pp. 136--143,
  December 2014.

\bibitem{wang2018compact}
Mingkai Wang, Yixin Li, Huanqing Zou, Mingzhi Peng, and Guangli Yang,
\newblock ``{Compact MIMO Antenna for 5G Portable Device Using Simple
  Neutralization Line Structures},''
\newblock in {\em 2018 IEEE International Symposium on Antennas and Propagation
  \& USNC/URSI National Radio Science Meeting}. IEEE, 2018, pp. 37--38.

\bibitem{giordani2016comparative}
Marco Giordani, Marco Mezzavilla, C~Nicolas Barati, Sundeep Rangan, and Michele
  Zorzi,
\newblock ``{Comparative analysis of Initial Access techniques in 5G mmWave
  cellular networks},''
\newblock in {\em {2016 Annual Conference on Information Science and Systems
  (CISS)}}. IEEE, 2016, pp. 268--273.

\bibitem{wei2017exhaustive}
Lili Wei, Qian Li, and Geng Wu,
\newblock ``{Exhaustive, Iterative and Hybrid Initial Access techniques in
  mmWave communications},''
\newblock in {\em {2017 IEEE Wireless Communications and Networking Conference
  (WCNC)}}. IEEE, 2017, pp. 1--6.

\bibitem{parada2017cell}
Raul Parada and Michele Zorzi,
\newblock ``{Cell discovery based on historical user's location in mmWave
  5G},''
\newblock in {\em {European Wireless 2017; 23th European Wireless Conference}}.
  VDE, 2017, pp. 1--6.

\bibitem{7343692}
A.~{Maltsev}, I.~{Bolotin}, A.~{Pudeyev}, G.~{Morozov}, and A.~{Davydov},
\newblock ``{Performance evaluation of the isolated mmWave small cell},''
\newblock in {\em 2015 IEEE 26th Annual International Symposium on Personal,
  Indoor, and Mobile Radio Communications (PIMRC)}, Aug 2015, pp. 2354--2358.

\bibitem{perera2019initialThesis}
Indika Perera,
\newblock ``{AN INITIAL ACCESS OPTIMIZATION ALGORITHM FOR MILLIMETRE WAVE 5G NR
  NETWORKS},''
\newblock {\em http://urn.fi/URN:NBN:fi:oulu-201910223006}, 2019.

\bibitem{perera2019initial}
A~Indika Perera, KB~Manosha, Nandana Rajatheva, and Matti Latva-aho,
\newblock ``{An Initial Access Optimization Algorithm for millimeter Wave 5G NR
  Networks},''
\newblock {\em arXiv preprint arXiv:1910.14359}, 2019.

\bibitem{ETSITS1383001}
{ETSI},
\newblock ``{5G; NR; Overall description; Stage-2},''
\newblock Technical Specification (TS) 138.300 v 15.3.1, {European
  Telecommunications Standards Institute (ETSI)}, 10 2018.

\bibitem{dahlman20185g}
{Dahlman, Erik and Parkvall, Stefan and Skold, Johan},
\newblock {\em {5G NR: The next generation wireless access technology}},
\newblock Academic Press, 2018.

\end{thebibliography}

\end{document}